\documentclass[aps,prb,superscriptaddress,ongbibliography,amsmath,amssymb,amsfonts,twocolumn]{revtex4-2}

\usepackage{graphicx}
\usepackage{dcolumn}
\usepackage{bm}
\usepackage{amsmath,amssymb}
\usepackage{microtype}
\usepackage{array}
\usepackage[usenames,dvipsnames]{xcolor}
\definecolor{nblue}{rgb}{0.0, 0.0, 1.0}
\definecolor{blue}{rgb}{0.0, 0.0, 1.0}
\definecolor{magenta}{rgb}{0.79, 0.08, 0.48}
\usepackage[colorlinks,linkcolor=nblue,urlcolor=blue,citecolor=magenta,plainpages=false,pdfpagelabels,breaklinks]{hyperref}
\usepackage[normalem]{ulem}

\renewcommand{\figurename}{Fig.}
\renewcommand{\tablename}{Table}
\makeatletter\renewcommand{\fnum@figure}[1]{\textbf{\figurename~\thefigure\,\textbar\,}}\makeatother
\makeatletter\renewcommand{\fnum@table}[1]{\tablename~\thetable\,\textbar\,}\makeatother

\begin{document}

\title{Flat Topological Nodal Lines in Heavy-Fermion Compound CeCoGe$_3$}

\author{Yuting Wang}
\affiliation{School of Science, Southwest University of Science and Technology, Mianyang 621010, China} 

\author{Weikang Wu$^*$}
\email[Corresponding author:\\]{weikang\_wu@sdu.edu.cn}
\affiliation{Key Laboratory for Liquid-Solid Structural Evolution and Processing of Materials (Ministry of Education), Shandong University, Jinan 250061, China}
\affiliation{Suzhou Research Institute of Shandong University, Suzhou, Jiangsu, 215123 China}

\author{Jianzhou Zhao$^\dagger$}
\email[Corresponding author:\\]{jzzhao@tju.edu.cn}
\affiliation{Center for Joint Quantum Studies and Department of Physics, School of Science, Tianjin University, 135 Yaguan Road, Tianjin 300350, China}


\begin{abstract}

The interplay between strong electronic correlations, unconventional superconductivity, and symmetry-protected topology provides a fertile ground for discovering exotic quantum states.
In this work, we investigate the correlated electronic structure and topological properties of the heavy fermion material CeCoGe$_3$ using density functional theory combined with dynamical mean-field theory calculations.
Our results reveal a crossover from high temperature incoherent states to low temperature coherent heavy quasiparticles, accompanied by a mass enhancement of $m^*/m_{\text{DFT}}\sim 52.6$ at $T=25$ K.
The interplay between electronic correlation, spin-orbit coupling and the noncentrosymmetric $I4mm$ crystal symmetry stabilize flat topological nodal lines within 10 meV of the Fermi level, which could contribute a significant density of states.
The proximity of topological nodal lines to the Fermi surface suggests a potential role in mediating pressure induced unconventional superconductivity.
Our work establishes CeCoGe$_3$ as a prototype topological nodal line Kondo semimetal. The coexistence of strong correlation,  non-trivial band topology and superconductivity indicate CeCoGe$_3$ as a potential candidate for realizing topological superconductivity.
\end{abstract}

\maketitle

\section*{Introduction}
Topological semimetals (TSMs) represent a class of quantum materials characterized by symmetry-protected band crossings near the Fermi level, which give rise to exotic quasiparticle excitations~\cite{armitage_weyl_2018,fang_multi-weyl_2012,bradlyn_beyond_2016,zhu_triple_2016,yu_encyclopedia_2022} and unconventional transport or optical responses~\cite{zyuzin_topological_2012,son_chiral_2013,huang_observation_2015,ashby_magneto-optical_2013,de_juan_quantized_2017}. Based on the degeneracy and spatial distribution of band crossing points (BCPs), TSMs can be classified into nodal point semimetals, such as Weyl semimetals (WSMs) and Dirac semimetals (DSMs), and nodal line semimetals (NLSs). In WSMs, doubly degenerate BCPs with non-zero Chern numbers emerge, requiring the breaking of either time-reversal or inversion symmetry.
Notable examples include pyrochlore iridates (Y$_2$Ir$_2$O$_7$~\cite{wan_topological_2011}) and transition metal monophosphides (TaAs~\cite{weng_weyl_2015}). In contrast, DSMs such as Na$_3$Bi and Cd$_3$As$_2$ preserve both time-reversal and inversion symmetries, where Weyl points of opposite chirality merge into fourfold-degenerate Dirac points under rotational symmetry protection~\cite{wang_dirac_2012,wang_three-dimensional_2013,liu_discovery_2014,neupane_observation_2014,liu_stable_2014}.
For NLSs~\cite{fang_topological_2016}, BCPs form closed loops in the Brillouin zone (BZ), classified as either ``accidental'' (removable via symmetry-preserving band deformation) or ``essential'' (rigidly tied to crystalline symmetries)~\cite{li_nonsymmorphic-symmetry-protected_2018}. The interplay between strong electron correlations and topology further enriches the landscape, leading to proposals of topological Mott insulators~\cite{raghu_topological_2008,chen_realization_2021,trivedi_topological_2021}, topological Kondo phases~\cite{dzero_topological_2010,lu_correlated_2013,deng_plutonium_2013,cao_trivial_2020,liu_discovery_2024,hu_ceco_2p_2_2024}, and topological superconductors~\cite{qi_topological_2011}.

\begin{figure}[!ht]
	\centering
	\vspace{-1ex}%
	\includegraphics[width=0.46\textwidth]{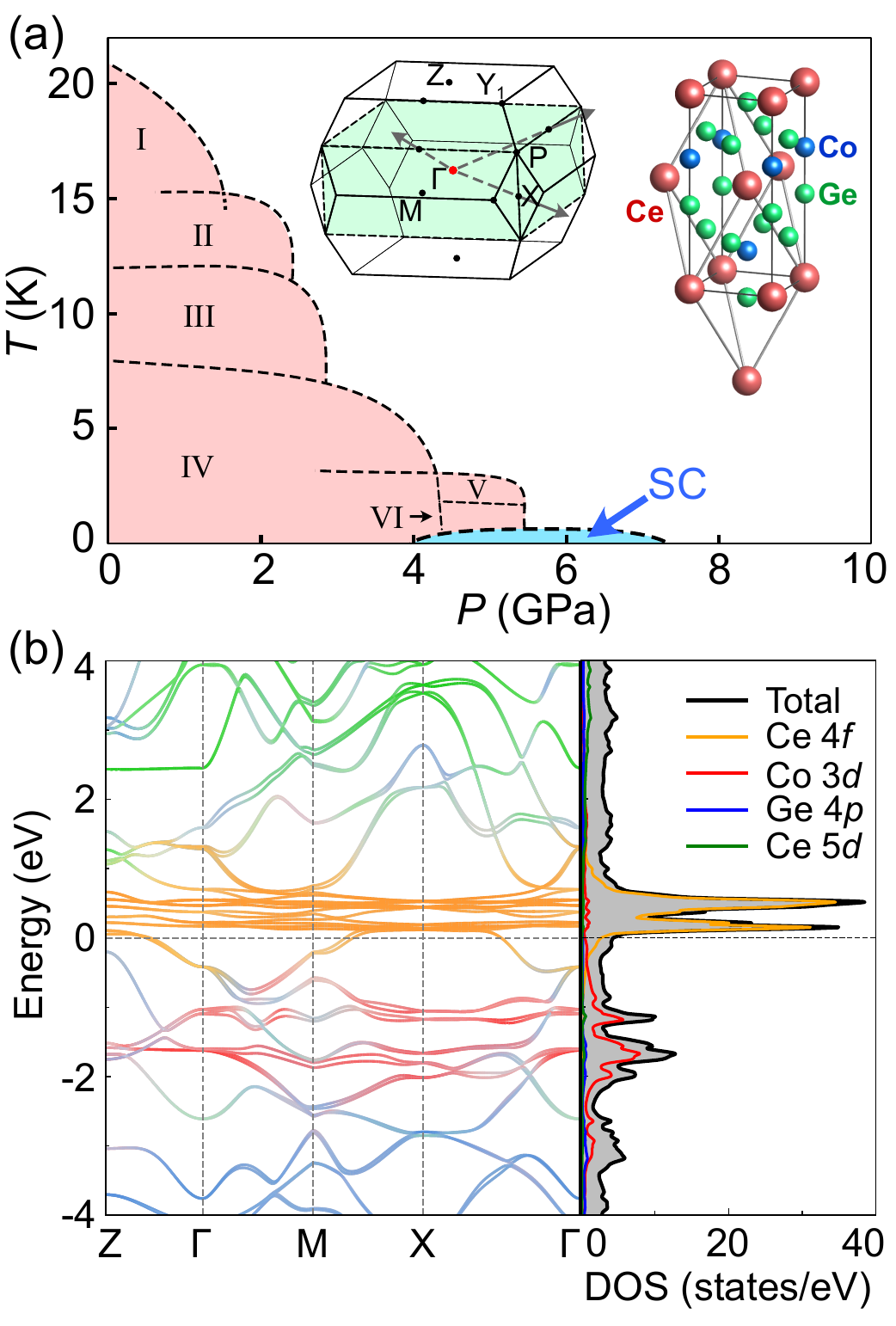}
	\caption{
		(a) Pressure-temperature phase diagram of CeCoGe$_3$ adapted from Ref.~\cite{knebel_high_2009}, illustrating six magnetic phases (in red) and the emergence of superconductivity under pressure (in blue). Inset: Body-centered tetragonal crystal structure of CeCoGe$_3$ with space group $I4mm$ and its Brillouin zone.
		(b) DFT-calculated band structure and DOS with orbital projected contribution: Co-3$d$ (red), Ge-4$p$ (blue), Ce-5$d$ (green) and Ce-4$f$ (orange). The crystal structure was illustrated using the {\sc VESTA} software~\cite{momma_vesta_2008}.
	}\label{fig:fig1}
\end{figure}

Heavy-fermion (HF) systems, as prototypical strongly electronic correlation materials, give rise to remarkable emergent phenomena, including heavy quasiparticle formation below the Kondo temperature ($T_K$), quantum criticality, and unconventional superconductivity. Recent studies have revealed that nontrivial topology can coexist with strong correlations in HF compounds, exemplified by the discovery of Weyl-Kondo semimetals~\cite{lai_weylkondo_2018,guo_evidence_2018}. 
Among Ce-based HF materials, CeCoGe$_3$ distinguishes itself through its unique structural and electronic properties. It crystallizes in a non-centrosymmetric BaNiSn$_3$-type tetragonal structure with $I4mm$ space group, as shown in the inset of Fig.~\ref{fig:fig1}(a).
It undergoes multiple magnetic transitions($T_{N1}=21$ K, $T_{N2}=12$ K, $T_{N3}=8$ K) at ambient pressure~\cite{pecharsky_unusual_1993,thamizhavel_unique_2005,smidman_neutron_2013}.
High-resolution angle-resolved photoemission spectroscopy (ARPES) studies~\cite{li_photoemission_2023} have uncovered a delicate competition between Kondo screening and magnetic ordering, highlighting the coexistence of localized $4f$ moments and itinerant electrons.
Under hydrostatic pressure, CeCoGe$_3$ exhibits a rich phase diagram (Fig.~\ref{fig:fig1} (a)) with six magnetic states and pressure-induced superconductivity ($T_C\sim 0.69$ K at 6.5 GPa)~\cite{kawai_magnetic_2008,knebel_high_2009}.
Despite extensive investigations into its magnetic and superconducting behaviors, the interplay between noncentrosymmetric crystal symmetry, strong correlations, and potential topological states in CeCoGe$_3$ remains largely unexplored.

In this work, we investigated the correlated electronic structure and unconventional topological properties of paramagnetic CeCoGe$_3$ using the density functional theory (DFT) combined with dynamical mean-field theory (DMFT) method~\cite{georges_dynamical_1996,kotliar_electronic_2006}.
Our results reveal that CeCoGe$_3$ exists strongly renormalized symmetry-protected nodal lines near the Fermi level in the low temperature paramagnetic phase. These findings not only provide a microscopic understanding of its HF behavior ($m^*/m_{DFT}\sim 52.6$) but also establish CeCoGe$_3$ as a candidate for realizing topological superconductivity. The proximity of nodal lines to the Fermi surface suggests their potential role in shaping unconventional superconducting pairing under pressure, offering new insights into the collaboration between correlation-driven topology and quantum criticality in noncentrosymmetric HF systems.

\section*{RESULTS}

\begin{figure}[!ht]
	\centering
	\vspace{-1ex}%
	\includegraphics[width=0.48\textwidth]{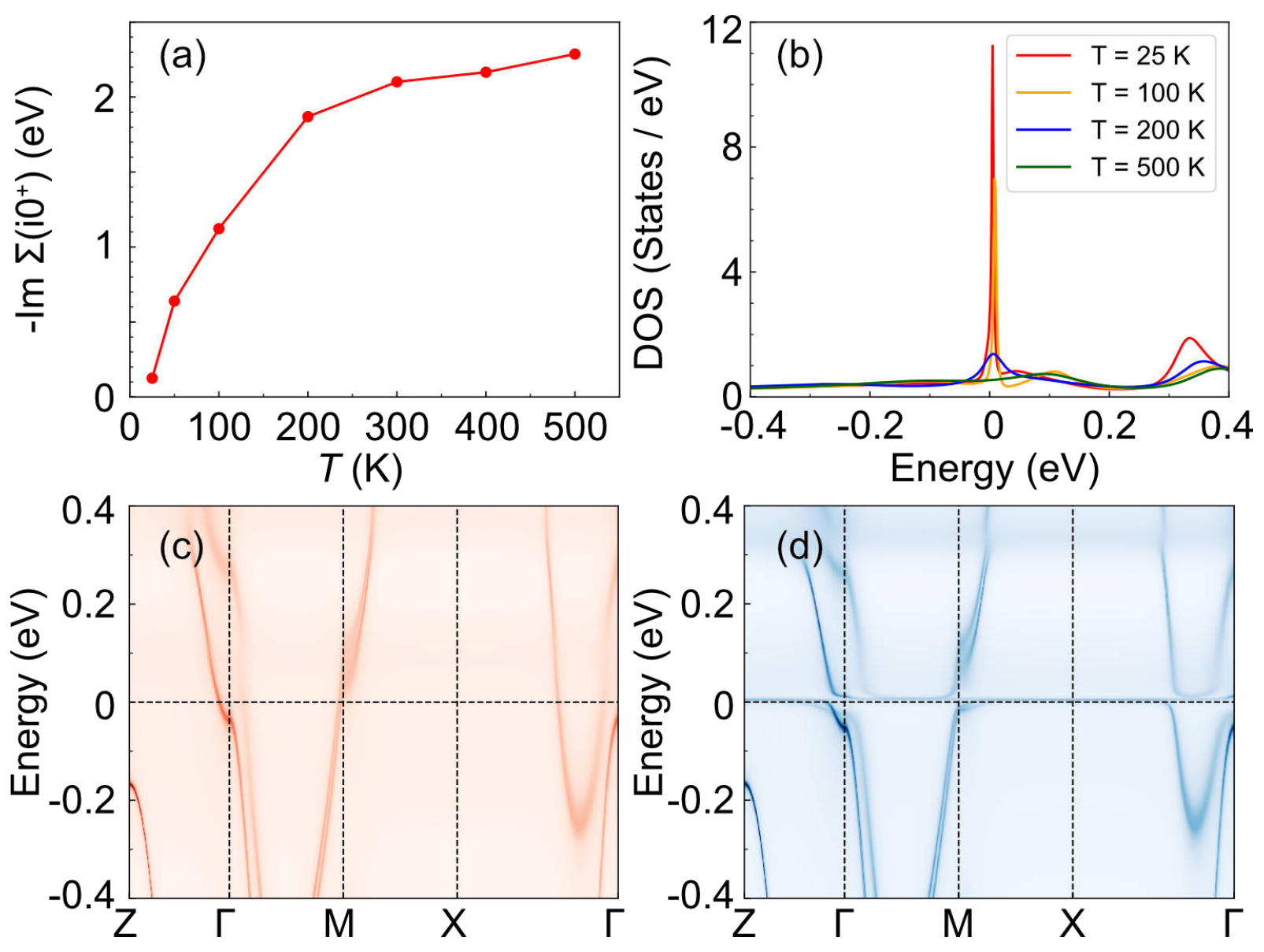}
	\caption{
		Temperature evolution of correlation-driven electronic strcture. (a) Imaginary part of the self-energy $-\mathrm{Im}\Sigma(\mathrm{i}0^+)$ for Ce-$f_{5/2}$ orbital as a function of temperature. (b) 4$f_{5/2}$-projected DOS at Fermi level as a function of temperature; (c) Momentum-resolved spectral function $\mathcal{A}(k,\omega)$ at $T=500$ K; (d) $\mathcal{A}(k,\omega)$ at $T=25$ K.
	}\label{fig:fig2}
\end{figure}

\subsection*{Correlation-driven electronic reconstruction}
The DFT band structure and density of states (DOS) of CeCoGe$_3$ is shown in Fig.\ref{fig:fig1} (b). 
One observes that the bands close to the Fermi level is mainly contributed by Ce-4$f$, Ce-5$d$, Co-3$d$ and Ge-4$p$ orbitals (seperated figures for each orbital are shown in Fig. S2 of supplemental materials). 
The occupation numbers are 1.03 and 0.66 for Ce 4$f$ and 5$d$ electrons respectively.
The strong spin-orbit coupling split the $4f$ orbitals into $4f_{5/2}$ and $4f_{7/2}$ states, and the bandwidth is about 0.7 eV.
The $4f_{5/2}$ states located near the Fermi level, hybridize with dispersive Co-$3d$ and Ge-$4p$ bands (shown in red and blue in Fig.~\ref{fig:fig1} (b)) forming multiple bands crossings.
However, conventional DFT calculation fail to capture the strong correlation effect of $4f$ electrons, resulting in an underestimation of the effective mass enhancement and failure to reproduce the Kondo effect observed in specific heat~\cite{pecharsky_unusual_1993,thamizhavel_unique_2005} and ARPES measurements~\cite{li_photoemission_2023}.
Therefore, a DFT+DMFT investigation is required.

Fig.~\ref{fig:fig2} (a) shows the temperature evolution of the imaginary part of the Matsubara self-energy $-\mathrm{Im}\Sigma(\mathrm{i}0^+)$.
At $T=500$ K, which is substantially higher than the Kondo temperature $T_K$, $-\mathrm{Im}\Sigma(\mathrm{i}0^+)$ is approximately 2.29 eV indicating a Mott-like incoherence behavior. This value decreases with temperature, showing a fast decay around $T=50$ K that signals the onset of coherence. 
Our self-energy evolution with temperatures is consistent with the temperature dependent ARPES experimental results, which indicate that the Kondo resonance peak starts to emerge around 80 K~\cite{li_photoemission_2023}.
It eventually reaches 0.125 eV at $T=25$ K, which remains above antiferromagnetic ordering temperature $T_{N1}$. 
The static local spin susceptibility $\chi$ vs $T$ exhibit a Curie-Weiss behavior in our calculation temperature range (see Fig. S3 in supplemental material).
The effective mass enhancement by correlations is defined by $m^*/m_{\text{DFT}} = 1/\mathcal{Z}$, where $\mathcal{Z}$ is the quasi-particle weight defined as 
\begin{equation}
	\mathcal{Z} = \left[1-\frac{\partial\text{Im}\Sigma(\mathrm{i}\omega)}{\partial\omega}\bigg|_{\omega\rightarrow 0^+}\right]^{-1}
\end{equation}
To avoid large error-bar in analytic continuation, we directly obtain $\mathcal{Z}$ from the polynormial fit to self-energies on the first seven Matsubara frequencies. The quasi-particle weight $\mathcal{Z}$ from our DMFT calculation is approximately 0.019 at $T=25$ K, corresponding to an the effective mass enhancement $m^*/m_{\text{DFT}}=52.6$. This value aligns with the heavy fermion character in CeCoGe$_3$ revealed by the de Haas-van Alphen measurements ($m^* \sim 30 m_e$)~\cite{sheikin_high-field_2011} and the specific heat coefficient $\gamma = 111$ mJ/(K$^2$\ mol)~\cite{pecharsky_unusual_1993}. The effective mass enhancement in CeCoGe$_3$ is much larger than topological HF semimetals studied previously~\cite{guo_evidence_2018,xu_heavy_2017,guo_possible_2017,settai_observation_1994}.

The correlation-induced mass enhancement is further confirmed by the temperature evolution of the $f$-projected DOS at the Fermi level, as shown in Fig.~\ref{fig:fig2} (b).
The DOS is nearly flat around $E_f$ at $T=500$ K, with no quasiparticle peak observed. As the temperature decreases, a quasiparticle peak emerges at $E_f$ at $T=200$ K, and grows systematically as the temperature decreases further, forming a very sharp peak at $T=25$ K.
The full width at half-maximum of spectral function $\Gamma$ at $T=25$ K is approximately 4 meV, from which the Kondo temperature can be estimated as $T_K = \Gamma / k_B = 46$ K. The value is consistent with the results obtained from the self-energy analysis.
The $4f$ occupation number remains at 1.05 from high temperature to low temperature in our DFT+DMFT calculations.

\begin{figure}[!ht]
	\centering
	\vspace{-1ex}%
	\includegraphics[width=0.49\textwidth]{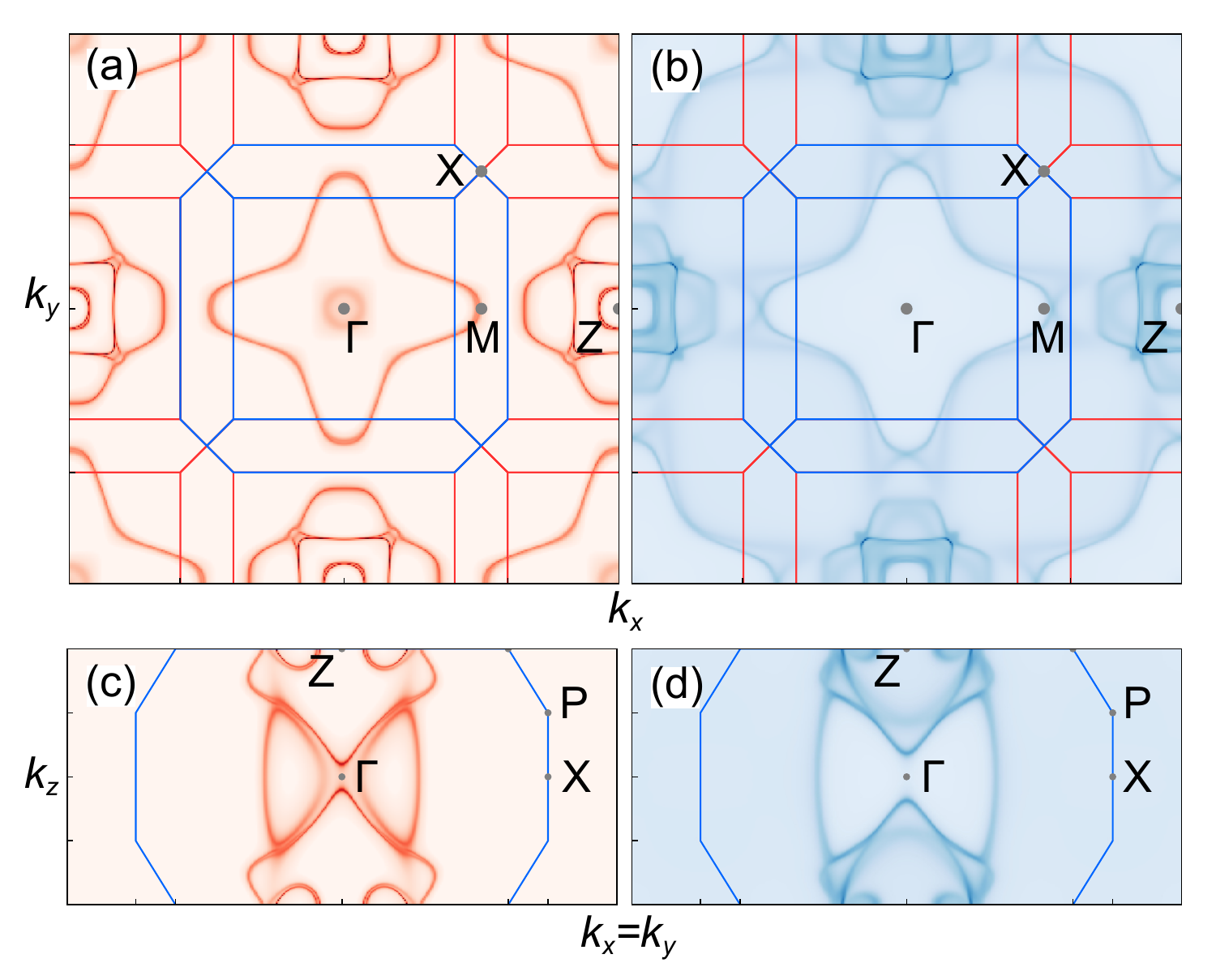}
	\caption{
		The Fermi surface evolution across the Kondo temperature. (a,c) The high temperature ($T=500$ K) Fermi surface in the (001) (a) and (110) (c) planes. (b,d) The low temperature ($T=25$ K) Fermi surface in the (001) (b) and (110) (d) planes.
	}\label{fig:fig4}
\end{figure}

\begin{figure*}[!ht]
	\centering
	\vspace{-1ex}%
	\includegraphics[width=0.9\textwidth]{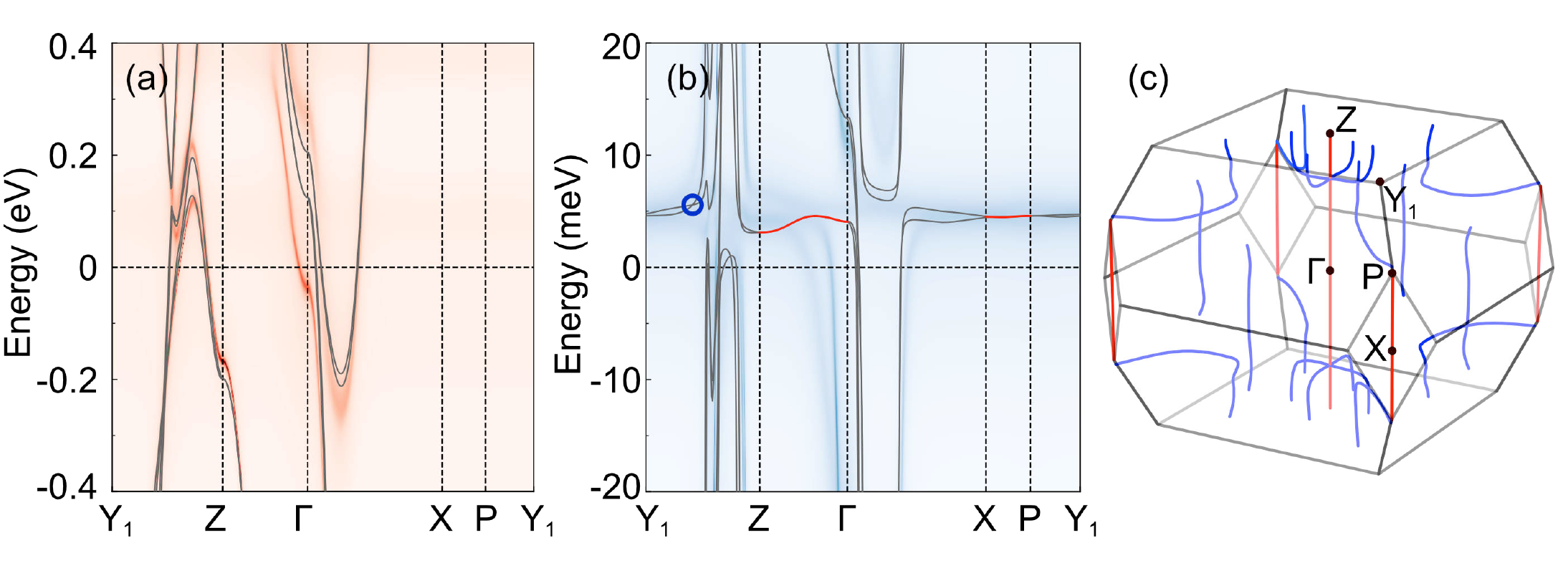}
	\caption{
		Effective quasiparticle band strcuture and topological nodal lines. (a) The high-temperature ($T=500$ K) band structure of the effective Hamiltonian, with spectral function $\mathcal{A}(k,\omega)$ overlaid (red). (b) The low-temperature ($T=25$ K) band structure. (c) The essential and accidental nodal lines in the BZ with red and blue colors respectively. The position of nodal lines are marked in (b) for references.
	}\label{fig:fig3}
\end{figure*}

The momentum-resolved spectrum $\mathcal{A}(k,\omega)$ at higher ($T=500$ K) and lower ($T=25$ K) temperature are depicted in Fig.~\ref{fig:fig2} (c) and (d). One could obviously observe the crossover from incoherent spectral function to coherent quasiparticle bands as the temperature decreases.
At high temperature ($T=500$ K), the spectral function in Fig.~\ref{fig:fig2} (c) exhibits broad coherent itinerant bands. Two van Hove singularities at $\Gamma$ point and $M$ point are observed, which are contributed by Ge-$4p$ and Co-$3d$ orbitals respectively (see more details about the orbital projected band structure with Ce-$4f$ as core levels in Fig. S4 of the supplemental material).
The $4f$ states are nearly invisible due to strong scattering and lack of coherence, which is aligned with the incoherent behavior of the self-energy at high temperature, indicating the suppressed quasiparticle formation.
At lower temperature, flat quasiparticle bands with $4f_{5/2}$ charater emerges at $E_f$ (Fig.~\ref{fig:fig2} (d)), signifying the establishment of Kondo coherence below $T_K$.
The observed flat bands with $4f$ characters are situated within 10 meV of the Fermi level, suggesting a potential contribution to novel transport properties and superconductivity. 
Furthermore, the intersection of these localized $4f$ bands with itinerant conduction bands near Fermi level results in band crossings.
Notably, the interplay between spin-orbit coupling and noncentrosymmetric crystal symmetry stabilizes these crossings, leading to the formation of topological nodal lines and Weyl semimetals, as discussed below.

The comparison of reconstructed Fermi surface at high temperature $T=500$ K and low temerature $T=25$ K further highlight correlation-driven topology in Fig.~\ref{fig:fig4}. 
Above $T_K$, the Fermi surface is mainly contributed by the itinerant conduction bands, with a large hole pocket at $\Gamma$ point from Ge-$4p$ and Co-$3d$ states and a small pocket from Ge-$4p$ states. 
Below $T_K$, the $4f$-derived flat bands intersect with the itinerant conduction bands near the Fermi level, forming hybridization gap around $\Gamma$ point, and eliminating the smaller pocket.

\subsection*{Topological nodal lines}
To gain a more comprehensive understanding of the topological properties of CeCoGe$_3$, we first constructed the tight-binding Hamiltonian $H_0$ from DFT using the Wannier functions~\cite{pizzi_wannier90_2020}, with the Ce-$4f$ orbitals treated as correlated and the Ce-$5d$, Co-$3d$ and Ge-$4s, 4p$ orbitals as uncorrelated basis functions.
To capture the low-energy heavy quasiparticle physics, we constructed the effective quasiparticle Hamiltonian $H_{QP}$ given by
\begin{equation}~\label{eqn:Hqp}
	H_{QP} = \sqrt{\mathcal{Z}} \left[ H_0 + \Re\Sigma(0) - \mu \right] \sqrt{\mathcal{Z}}
\end{equation}
where $\mu$ is the chemical potential. It is crucial to note that, $\sqrt{\mathcal{Z}}$ and $\Re\Sigma(0)$ are diagonal matrices in orbital space. The diagonal elements corresponding to the correlated Ce-4$f$ orbitals are the square root of renormalization factor ($\sqrt{\mathcal{Z}}$) and the real part of the self-energy at zero frequency ($\Re\Sigma(\omega=0)$) obtained from our DFT+DMFT calculations, while those for the uncorrelated orbitals are 1 and 0.
The quasiparticle Hamiltonian defined in Eqn.~\ref{eqn:Hqp} only holds in the low temperature coherent regime where the imaginary part of self-energy at zero frequecy is negligible. Our DMFT results at $T=25$ K approximately satisfy this condition. To study the coherent band structure at higher temperature ($T=500$ K), we shift the localized Ce-$4f$ levels away from the Fermi level according to the real part of self-energy $\Re\Sigma(0)$ at $T=500$ K, while neglecting the renormalization factor $\mathcal{Z}$.
The band structure of the effective Hamiltonian at higher ($T=500$ K) and lower ($T=25$ K) temperature are shown in Fig.~\ref{fig:fig3} (a) and (b) respectively, which capture the key features of $\mathcal{A}(k,\omega)$ spectra as shown in background.
As discussed above, the main features of the band structure at low temperature are the flat bands near the Fermi level crossing with the itinerant conduction bands.
Using the effective quasiparticle Hamiltonian, we confirmed the existence of Weyl points in the lower temperature phase. 
However, we mainly focus on the flat nodal lines induced by the electronic correlation here. Because they can contribute a large DOS and potentially host unconventional pairing channels, which have been suggested either by the highly renormalized bands in strongly correlation systems~\cite{checkelsky_flat_2024,chen_coexistence_2024,cao_flat_2024,xu_unraveling_2024,baglo_fermi_2022} or by van Hove sigularity in Kagome lattice~\cite{yu_chiral_2012,wu_nature_2021,wilson_av3sb5_2024}.

Through scanning the whole BZ~\cite{wu_wanniertools_2018}, we found two types of nodal lines whose locations are exhibited in Fig.~\ref{fig:fig3} (c). One is the essential Weyl nodal line with double degeneracy along $\Gamma$-$Z$ and $X$-$P$ (red lines in Fig.~\ref{fig:fig3} (c)). Their presence is solely determined by space group symmetries and do not need band inversion. Specifically, the degeneracy is due to the anticommutation relation between the twofold rotation $C_{2z}$ and a vertical mirror $M_v$ on these paths in the presence of spin-orbit coupling. Here, $v$ indicates the normal direction of the mirror plane. For $\Gamma$-$Z$, it is $\{C_{2z}, M_{x/y}\} = 0$, while $\{C_{2z}, M_{110/1\bar{1}0}\} = 0$ applies for the $X$-$P$ path. The anticommutation relation originates from the anticommutativity between two spin rotations, e.g. $\{\sigma_z, \sigma_{x/y}\} = 0$ under SOC. Consequently, for an eigenstate $|c_{2z}\rangle$ of $C_{2z}$ with the $C_{2z}$ eigenvalue $c_{2z}$, the following relation folds:
\begin{equation}
	C_{2z}(M_v|c_{2z}\rangle) = - c_{2z}(M_v|c_{2z}\rangle)
\end{equation}
This illustrates that $|c_{2z}\rangle$ and $M_v|c_{2z}\rangle$ are distinct states but degenerate at the same energy. Off $\Gamma$-$Z$ or $X$-$P$, the anticommutation relation do not hold. Thus, this ensures a Weyl nodal line along $\Gamma$-$Z$ or $X$-$P$.

The other nodal lines sit on four vertical mirror planes, $M_{x/y}$ and $M_{110/1\bar{1}0}$, as shown in blue lines in Fig.~\ref{fig:fig3}(c). These nodal lines are protected by the four vertical mirror symmetries and are formed by the crossing of two bands with opposite mirror eigenvalues, which are $\pm i$ in the presence of SOC. They are of accidental type, because their formation requires band inversion and they would be removed without breaking symmetries. 
Both types of nodal lines are topological. Since they all sit on mirror planes, the mirror symmetry can dictate a $\mathbb{Z}$-topological classification with the topological invariant $\xi$ defined as $\xi=N_1-N_2$~\cite{fang_topological_2016}. Here, $N_1$ and $N_2$ denote the number of bands with the +i mirror eigenvalue at two k-points on two sides of a nodal line. A nonzero $\xi$ indicates a topological nodal line. We calculate $\xi$ for two types of nodal lines, which both gives $|\xi|=1$ and thus they are all topologically protected.
It is noted that, previous work revealed a set of nodal lines extending from a Dirac point in CeCoGe$_3$ using the LDA + Gutzwiller method~\cite{ivanov_renormalized_2021}. These nodal lines are also formed by band inversion and thus of accidental type. They are protected by mirror symmetries and robust against perturbations.

It is critical to emphasize that no structural phase transition has been experimentally observed in CeCoGe$_3$ across the temperature range from $T=500$ K to $T=25$ K. This implies that the essential nodal lines protected by crystal symmetry at low temperature, persist in the high temperature regime. However, their visibility and physical relevance at high temperature are suppressed by two key factors. First, the significant energy dispersion of the itinerant bands shift the nodal lines far from the Fermi level. Second, the strong scattering from incoherent $4f$ states smear out sharp features. Consequently, these symmetry protected flat nodal lines are spectroscopically unresolved at high temperature , and unlikely to dominate unconventional transport properties.

\subsection*{Topological properties under pressure}
Below $T_{N1}=21$~K, CeCoGe$_3$ undergoes succesive antiferromagnetic (AFM) phase transitions accompanied by long-range magnetic ordering. 
Notably, hydrostatic pressure suppresses the AFM order while preserving the crystal structure (Fig.~\ref{fig:fig1} (a)), as evidensed by the absence of structural phase transition under pressure.
This implies that the symmetry protected topological nodal lines, identified in our work, persist in the paramagnetic phase under pressure. 
To connect the topological properties with the pressure-induced superconductivity, we investigate the correlated electronic structure at $P=6.5$ GPa, where the maximum $T_c$ is experimentally observed.

\begin{figure}[!ht]
	\centering
	\vspace{-1ex}%
	\includegraphics[width=0.49\textwidth]{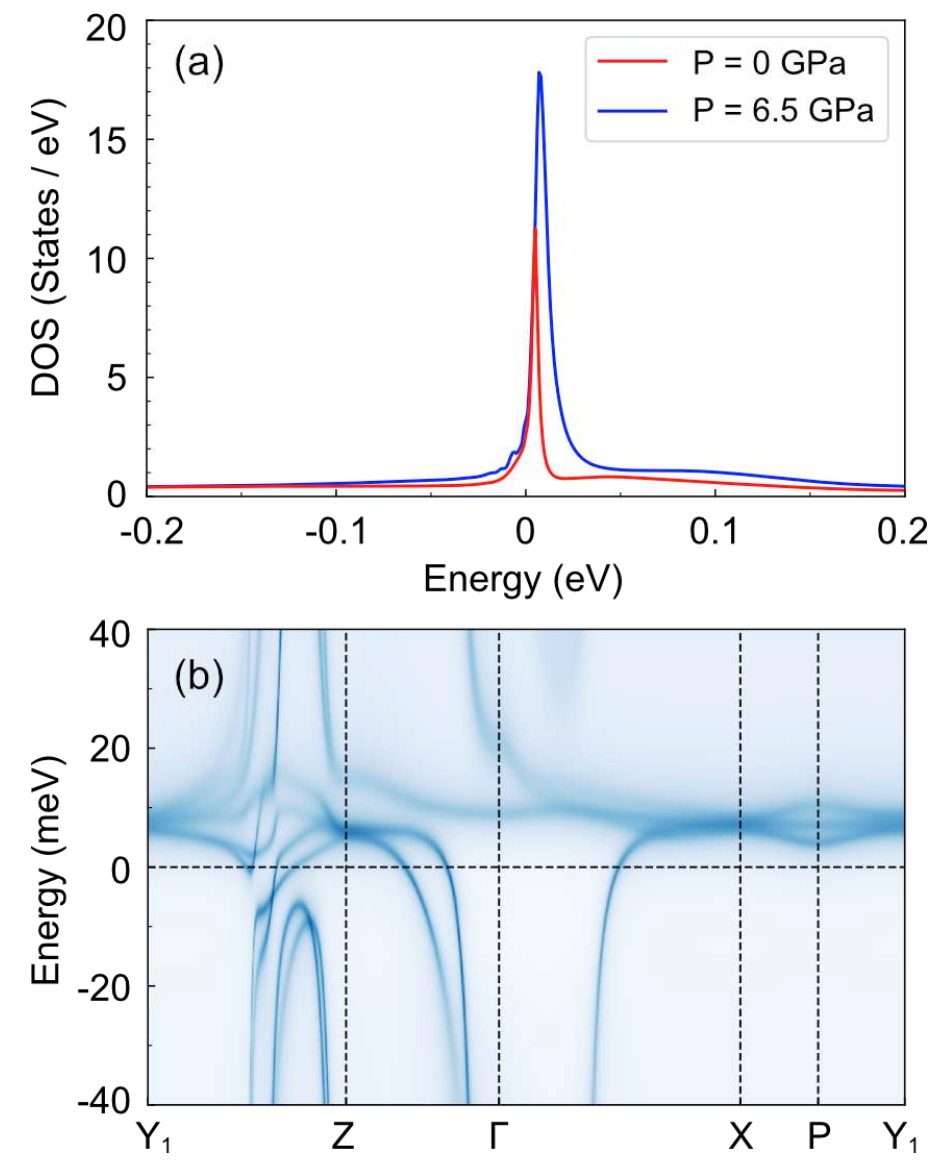}
	\caption{
		Calculated electronic structure of CeCoGe$_3$ at $P=6.5$ GPa. (a) The Ce-$4f$ projected DOS (blue line), comparing with  Ce-4$f$ DOS at the ambient pressure shown by red line. (b) Momentum-resolved spectral function $\mathcal{A}(k,\omega)$.
	}\label{fig:fig5}
\end{figure}

In contrast to the substantial effective mass enhancement at ambient pressure ($m^*/m_{\text{DFT}}=52.6$), the value at $P=6.5$ GPa and $T=25$ K is significantly reduced to approximately 22.6.
This reduction is consistent with the expectation that the enhanced kinetic energy of electrons under pressure leads to a lower effective mass. 
Despite this change in mass enhancement, the $4f$ occupation number of 1.04 at 6.5 GPa remains similar to the value at ambient pressure. 
From the Ce-$4f$ projected DOS at $P=6.5$ GPa in Fig.~\ref{fig:fig5} (a), one observe that the $4f$ quasiparticle peak is broader and higher compared to ambient pressure, indicating a larger bandwidth and a smaller effective mass.

Fig.~\ref{fig:fig5} (b) presents the momentum resolved spectral function at $P=6.5$ GPa. 
Notably, it exhibits similarities to the spectrum observed at ambient pressure.
More importantly, the characteristic flat nodal lines along Z - $\Gamma$ and X-P directions are still observed around the Fermi level (see also the quasiparticle band structure in Fig. S5 of supplemental materials), as expected from our symmetry analysis. The significant decrease in $-\mathrm{Im}\Sigma(\mathrm{i}0^+)$ from 125 meV at ambient pressure to 9.5 meV at $P=6.5$ GPa, indicate a significantly enhancement of the quasiparticle lifetime. 
Consequently, this enhanced coherence yields a much sharper spectral function at high pressure.

\section*{DISCUSSION}
We have revealed the presence of flat topological nodal lines due to the interplay of strong electronic correlations, spin-orbit coupling, and noncentrosymmetric crystal symmetry in CeCoGe$_3$ using \textit{ab initio} DFT+DMFT calculations.
These flat nodal lines reside within $\pm 10$ meV of the Fermi level, generating a large density of states. This enhanced DOS can amplify electronic correlations, potentially driving the exotic transport properties and pressure induced superconductivity observed in CeCoGe$_3$~\cite{liu_nodal_2025}. Experimentally, it exhibit a dome-shaped superconducting phase under pressure~\cite{eom_magnetic_2006,eom_non-fermi_2000}, accompanied by non-Fermi liquid behavior near the quantum critical point.
The strong SOC in CeCoGe$_3$ further enriches this interplay. The momentum-locked spin textures around the nodal lines can favor unconventional pairing symmetry~\cite{ivanov_renormalized_2021}, consistent with the anomalously high upper critical field $H_{c2}$ exceeding the Pauli limit~\cite{measson_huge_2010}.

Our results suggests a interplay between correlation driven topology, quantum critical fluctuations and topological superconductivity. We propose CeCoGe$_3$ as a potential candidate for realizing topological superconductivity~\cite{norman_hunds_1994, hoshino_superconductivity_2015, wu_nodal_2023}.
Further experimental studies under pressure, such as low temperature scanning tunneling microscopy, muon-spin spectroscopy and pressure dependent ARPES  measurements, are needed to confirm the superconductivity gap symmetry and the role of nodal lines in mediating unconventional pairing in CeCoGe$_3$.

\section*{Methods} 
We perform fully charge self-consistent DFT+DMFT calculations using the {\sc edmftf} package~\cite{Haule.2010}, based on the full-potential linear augmented plane-wave method implemented in the {\sc wien2k} code~\cite{Blaha.2020}.
The $k$-point mesh for the Brillouin zone integration is $12\times 12\times 12$, and the plane wave cut-off $K_{max}$ is given by $R_{MT}\times K_{max}=7.5$. We employed the generalized gradient approximation (GGA) with the Perdew-Burke-Ernzerhof (PBE) realization~\cite{Perdew.1996} as exchange-correlation functional. The atomic sphere $R_{MT}$ are 2.60, 2.25, 2.14 a.u. for Ce, Co and Ge, respectively.
We use projectors with energy window from $-10$ to 10 eV relative to the Fermi level to construct Ce-$4f$ local orbitals.
We choose Coulomb interaction parameter $U=6.0$ eV and Hund's exchange parameter $J_H=0.7$ eV in this work, which are typical values as used previously for Cerium~\cite{zolfl_spectral_2001,amadon__2006,lanata__2013}. The impurity problem is solved by the hybridization expansion version of the continuous time quantum Monte Carlo (CTQMC) solver~\cite{Gull.2011}.
We choose an ``exact'' double counting scheme developed by Haule~\cite{Haule.201509p}, in which the Coulomb repulsion in real space is screened by the combination of Yukawa and dielectric functions. The self-energy on real frequency is obtained by the analytical continuation method of maximum entropy~\cite{Haule.2010}.
We have considered spin-orbit coupling in the calculation.
We adopt the experimental lattice parameters in the calculation~\cite{smidman_neutron_2013}.
To investigate the correlated electronic structure under superconducting pressure, we fitted the Birch-Murnaghan equation of state (EOS)~\cite{birch_finite_1947} based on the calculated energy-volume relationship. For each volume considered, we fully relaxed all lattice parameters and atomic positions.
The structural relaxation of CeCoGe$_3$ was performed within the framework of DFT+$U$ as implemented in the VASP package~\cite{kresse_efficient_1996}, and utilizing the approach proposed by Dudarev \textit{et al}.~\cite{dudarev_electron-energy-loss_1998}. An effective onsite Coulomb interaction parameter $U_{eff} = 2$ eV was chosen, as this value reproduces the experimental equilibrium volume.

\section*{Acknowledgments} 
We thank T. Shang and L. H. Hu for very helpful discussion.
J. Zhao acknowledges support by the National Key Laboratory of Shock Wave and Detonation Physics Fund (No. JCKYS2025212108) and the National Natural Science Foundation of China (No. 12474239). 
W. Wu acknowledges support by the Shandong Provincial Natural Science Foundation (No. ZR2023QA012), Basic Research Program of Jiangsu (No. BK20230241), the National Natural Science Fund for Excellent Young Scientists Fund Program (Overseas), and the program of Outstanding Young and Middle-aged Scholars of Shandong University.




\vspace{20pt}
\bibliography{cecoge3.bib}

\end{document}